\documentclass[reprint, aps, pre, superscriptaddress, nofootinbib,nobibnotes]{revtex4-2}
\usepackage[utf8]{inputenc}
\usepackage{wrapfig}
\usepackage{latexsym}
\usepackage{mathtools, amssymb, amsmath, nicefrac, mathrsfs}
\usepackage{hyperref}
\usepackage[dvipsnames]{xcolor}
\usepackage{soul}
\usepackage{t1enc}
\usepackage{array}%table formatting
\usepackage{placeins}

\usepackage{pifont}% http://ctan.org/pkg/pifont
\newcommand{\cmark}{\ding{51}}
\newcommand{\xmark}{\ding{55}}

\newcommand{\set}[1]{\mathcal{#1}}
\newcommand{\T}[1]{\textrm{#1}}

\newcommand{\V}[1]{{\bf #1}}

\newcommand{\avg}[1]{\left\langle #1\right\rangle}

\begin{document}

\title{Network dismantling by physical damage}

\author{Luka Blagojevi\'c}
\affiliation{Department of Network and Data Science, Central European University, Vienna, Austria}
\author{Ivan~Bonamassa}
\affiliation{Department of Network and Data Science, Central European University, Vienna, Austria}
\author{Márton~Pósfai}
\email{posfaim@ceu.edu}
\affiliation{Department of Network and Data Science, Central European University, Vienna, Austria} 

\date{\today}
\begin{abstract}
We explore the robustness of complex networks against physical damage.
We focus on spatially embedded network models and datasets where links are physical objects or physically transfer some quantity, which can be disrupted at any point along its trajectory.
To simulate physical damage, we tile the networks with boxes of equal size and sequentially damage them. 
By introducing an intersection graph to keep track of the links passing through tiles, we systematically analyze the connectivity of the network and explore how the physical layout and the topology of the network jointly affect its percolation threshold. 
We show that random layouts make networks extremely vulnerable to physical damage, driven by the presence of very elongated links, and that higher-dimensional embeddings further increase their vulnerability. 
We compare this picture against targeted physical damages, showing that it accelerates network dismantling and yields non-trivial geometric patterns. 
Finally, we apply our framework to several empirical networks, from airline networks to vascular systems and the brain, showing qualitative agreement with the theoretical predictions. 
\end{abstract}
\maketitle

\section{Introduction}

From the earliest days of complex network research, percolation processes were used to study the robustness of networks against external perturbations~\cite{albert2000error, callaway2000network, cohen2010complex}.
Since then, various variants were introduced, including percolation on spatially embedded networks~\cite{moukarzel2006percolation,li2011percolation,vaknin2020spreading,amit2023percolation}, interdependent percolation~\cite{leicht2009percolation,buldyrev2010catastrophic,brummitt2012suppressing}, explosive percolation~\cite{achlioptas2009explosive,d2019explosive}, and many others~\cite{lee2018recent,li2021percolation, schwarze2024structural,artime2024robustness}.
Here, we explore the robustness of spatial networks against physical damage. By physical damage, we mean the random or targeted removal of entire regions of the space, such that all nodes inside and links passing through are disrupted.
Hence, we focus on spatial networks where the links represent physical objects or the physical transfer of some quantity, such that the link can be disrupted at any point along its trajectory.
Note that recent literature defined physical networks as networks built from physical objects where volume exclusion plays an important role, i.e., the network is composed of tightly packed nodes and links~\cite{dehmamy2018structural,posfai2024impact,pete2024physical}.
Here, we model a broader class of systems that include networks such as the air traffic network, where volume exclusion may not be a significant factor, yet links can be disrupted along their path by physical perturbations such as storms, volcanic eruptions, or military conflicts.
 
To study the robustness of such physically embedded networks, we introduce and analyze a percolation process, where in each step we do not damage single elements of the network but entire regions of the embedding space, removing any links that pass through them.
Motivating examples of such spatially correlated failures include permanent damage, such as necrosis, of the brain or other biological tissues due to injury or diseases~\cite{aktas2007neuronal, taoufik2008ischemic}, or natural disasters like floods, earthquakes, storms and other localized attacks that disrupt entire regions of socio-technological infrastructure~\cite{duenas2007seismic, ouyang2014multi, argyroudis2015systemic, wang2019local}, or even the effect of targeted cyberattacks and military strikes that may render entire areas of a communication network (e.g.\ satellite or ground stations) inoperable.

To formally set up the problem, we represent a physically embedded network by a combinatorial network, $\set G$, and a physical layout, $\set P$, which provides the location and shape of the nodes and links.
Here, we focus on physical nodes represented by a point in space connected by straight or curved lines; however, our framework easily extends to alternative physical network representations, including nodes and links that occupy volume~\cite{dehmamy2018structural, posfai2024impact} or network-of-networks representations~\cite{bianconi2013superconductor, chepuri2023complex, pete2024physical}.

\begin{figure}[h]
	\centering
	\includegraphics[width=1.\columnwidth]{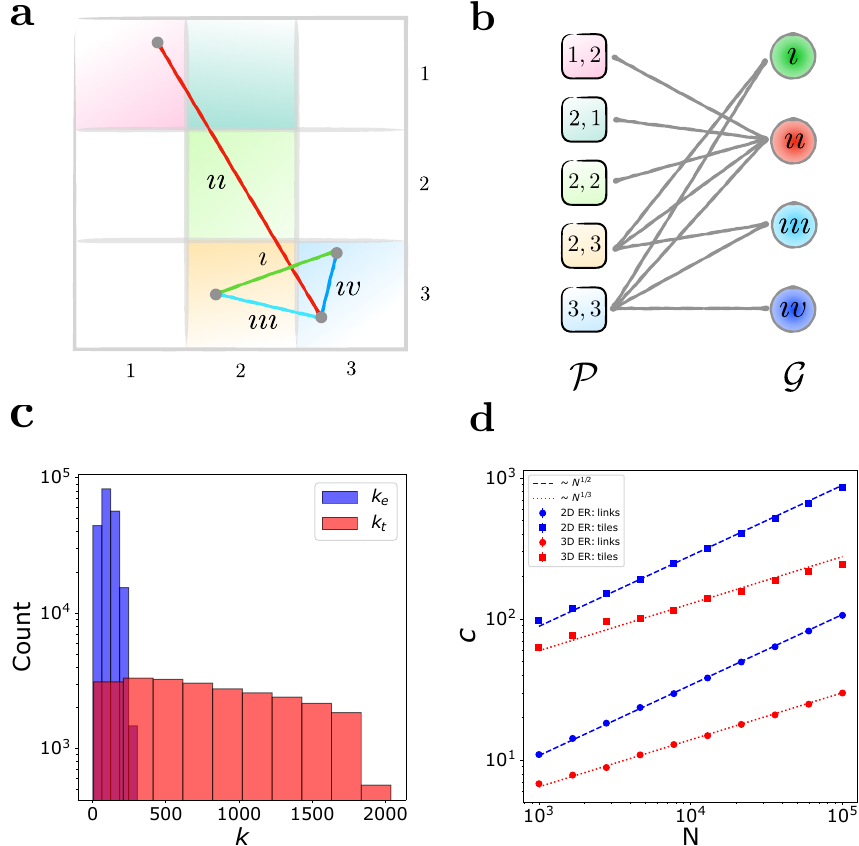}
	\caption{{ \bf Spatially correlated damage and the intersection graph.}
	(a)~We tile a physical network with $D$-dimensional square tiles. Damaging a tile $t$ removes all links $e$ intersecting $t$, for example, damaging tile $(3,3)$ (light blue) removes blue and red links from $\set G$.
	(b)~The bipartite intersection graph $\set I$ captures how tile damage of $\set P$ translates to link removal in $\set G$. Vertices on the left represent tiles and vertices on the right represent links; a tile and a link are connected if they intersect in the layout.
	(c)~The tile-side and link-side degree distribution of $\set I$ for a single randomly embedded ER network  ($N=10^{5},c=4,\rho =4 $).
	In random embeddings, links typically span the entire system, hence $\set I$ becomes dense.
	(d)~We increase the network size $N$ while keeping constant the expected number of nodes per tile, $\rho$. Both the tile-side average degree $c_\T t$ and link-side average degree $c_\T l$ scale as $\sim N^{-1/D}$.
	Networks are generated using parameters $c=4$ and $\rho = 4$; each marker represents an average of $20$ independent runs and the error bars represent the standard error of the mean. }
	\label{fig:setup}
\end{figure}

To capture the effects of physical damage, we first tile the $D$-dimensional space occupied by a physical network $\set P$ with $D$-dimensional cubes of side length $b$ (Fig.~\ref{fig:setup}a).
We then damage tiles sequentially and, when a tile $t$ is damaged, we remove from $\set G$ each link $e$ intersected by $t$.
As the tiles are removed, the combinatorial network $\set G$ undergoes a continuous percolation transition that is a mixture of site and bond percolation: 
If a node is inside a damaged tile, then all of its link are removed simultaneously, in effect removing the node.
On the other hand, if a link is traversing a damaged tile, then the link is removed but the nodes at its endpoints are not.
In this paper, we aim to understand how the physical layout $\set P$ and the structure of the combinatorial network $\set G$ jointly affect this critical transition.

The remainder of the paper is organized as follows.
In the next section, we introduce the intersection graph, an auxiliary graph that captures how tile damage translates to link removal.
The intersection graph and its randomizations serve as our main tool to numerically and analytically study the percolation transition.
In Secs.~\ref{sec:random} and \ref{sec:targeted}, we use randomly embedded model networks to analytically and numerically investigate the robustness of networks against random damage and targeted attacks.
Finally, in Sec.~\ref{sec:empirical}, we use randomizations of the intersection graph to probe how physical layout affects the critical transition in several empirical networks.

\section{Intersection graph}
\label{sec:intersect_graph}

We start by introducing the intersection graph, $\set I$, to concisely represent how tile damage translates to link removal.
First, we identify which physical link, $e$, intersects which tile, $t$, and we summarize these intersections as a bipartite graph, $\set I$: on one side of $\set I$ each vertex represents a tile $t$, on the other side each vertex represents a physical link $e$; a tile $t$ and a link $e$ is connected in $\set I$ if they intersect in the physical space (Fig.~\ref{fig:setup}b).

The intersection graph, $\set I$, encodes potential sources of heterogeneity of the physical layout that affect link removal:
(i)~long links intersect more tiles than short links, captured by the degree distribution $P_\T l(k)$ of the vertices that represent physical links in $\set I$ (Fig.~\ref{fig:setup}c); (ii)~in absence of periodic boundary conditions, centrally located tiles are expected to intersect more links than peripheral tiles, captured by the degree distribution $P_\T t(k)$ of the vertices that represent tiles (Fig.~\ref{fig:setup}c); (iii)~the spatial organization of $\set P$ is captured by structural correlations of $\set I$, for example, two physical links running approximately parallel to each other tend to intersect the same tiles, leading to an abundance of length-4 cycles in $\set I$.

Beyond its structure, an additional way $\set I$ carries information relevant to the percolation process is through its link-side vertex labels which connect $\set I$ and $\set G$.
For example, links attached to the same node $v$ must all intersect the tile $t$ that contains $v$; therefore, if $t$ is damaged, all links incident on $v$ are removed together from $\set G$.
Furthermore, because long links intersect many tiles, they become vulnerable to tile damage, and if such links tend to be important in $\set G$, the combinatorial network inherits their vulnerability.

To understand how the above heterogeneities and correlations of $\set I$ affect the robustness of physical networks, we randomize $\set I$ and compare the percolation transition guided by the original $\set I$ and its randomized version $\set I_\T{r}$.
Different randomizations allow us to probe the role of different features of the physical embedding. For instance, substituting $\set I$ with a bipartite configuration model with the same degree distributions $P_\T l(k)$ and $P_\T t(k)$, preserves link and tile heterogeneity but removes any additional structure (e.g., structure originating from parallel links or links sharing an endpoint) from $\set I$.
To remove link heterogeneity, we homogenize $\set I$ by setting the vertices representing links to have approximately the same degree equal to the average $c_\T l$.
In practice, this means that in the randomized $\set I_\T r$ the link-side vertices are a mixture of vertices with degree $\lceil  c_\T l \rceil$ and $\lfloor  c_\T l \rfloor$ in a way that ensures that the overall average degree remains $c_\T l$. 
Applying the same degree-homogenization to the tile vertices leads to four possible degree-preserving (DP) or homogenized (H) randomization protocols of $\set I$: (i)~tile and link degree-preserved (tDP-lDP), (ii)~tile degree-preserved and link homogenized (tDP-lH), (iii)~tile homogenized and link degree-preserved (tH-lDP), and (iv)~ tile and link homogenized (tH-lH).
Note that the randomized intersection graphs do not correspond to valid physical layouts because a link in $\set I_\T{r}$ generally intersects non-adjacent tiles.

The randomizations that preserve link-side degree in $\set I$ (tDP-lDP and tH-lDP) do not remove correlations between $\set G$ and $\set I$: if long links are important in $\set G$, then important links in $\set G$ tend to have high degree in $\set I$.
To study the effect of this correlation we introduce an additional randomization procedure called ``label shuffle'' (LS): we shuffle the vertex labels in $\set I$ and create a randomized intersection graph, $\set I_\T{LS}$, that is uncorrelated with $\set G$ but otherwise has the same structure as $\set I$.
This means that dismantling $\set G$ using $\set I_\T{LS}$ removes, on average, the same number of links as $\set I$, but the links are chosen randomly in $\set G$ independent of the original layout. 
We summarize in Table~\ref{tab:randomizations} the main features of the randomization protocols introduced above.
 
\begin{table}[h!]
\centering
\caption{{\bf Randomizations of the intersection graph.} Brief summary of properties preserved by each randomization; shared endpoint means that all links connected to a node $v$ intersect the tile that contains $v$.}
\label{tab:randomizations}
\newcolumntype{C}[1]{>{\centering\arraybackslash}p{#1}}
\begin{tabular}{p{1.5cm}*{4}{|C{1.3cm}}|C{.5cm}} 
 & tDP-lDP & tDP-lH & tH-lDP & tH-lH & LS \\\hline
$P_\T t(k)$    & \cmark & \cmark & \xmark & \xmark & \cmark \\\hline
$P_\T l(k)$    & \cmark & \xmark & \cmark & \xmark & \cmark \\\hline
$k(e)$-$\set G$ correlation    & \cmark & \xmark & \cmark & \xmark & \xmark \\\hline
Shared endpoint       & \xmark & \xmark & \xmark & \xmark & \xmark \\
\end{tabular}
\end{table}

\subsection{Link-side degree}

The degree $k(e)$ of a link $e$ in the intersection graph $\set I$ determines the vulnerability of $e$ to tile damage: the more tiles $e$ intersects, the more likely $e$ is removed. 
In this section, we describe the relation between the length of $e$ and $k(e)$.
We consider a physical network with $N$ nodes and $M$ links embedded in the unit $D$-dimensional cube such that the links are straight segments.
We tile the network with cubes of side length $b$, hence the intersection graph $\set I$ contains $n_\T{t}=b^{-D}$ vertices representing tiles and $n_\T{l}=M$ vertices representing links.
A link $e$ with endpoints $\V r=(r_1,r_2,\ldots,r_D)$ and $\V s=(s_1,s_2,\ldots,s_D)$ following a straight trajectory intersects a sequence of tiles $(t_1,t_2\ldots,t_{k(e)})$.
Whenever $e$ crosses from a tile $t_i$ to the next $t_{i+1}$, it punctures the shared face of $t_i$ and $t_{i+1}$, the probability of crossing at an edge or a corner is zero.
Therefore, link $e$ must cross at least $\lvert s_i - r_i \rvert/b$ tiles along each axis $i$, hence the approximate number of tiles intersected by $e$, i.e., its degree in $\set I$, is
\begin{equation}\label{eq:link_degree}
k(e) = 1+\sum_i \frac{\lvert s_i - r_i \rvert}{b},
\end{equation}
where the plus one corresponds to the starting tile.
Note that it is not the Euclidean but the Manhattan distance of the endpoints of $e$ that determines $k(e)$. 
For example, in a randomly embedded network, the typical link length $\ell$ is on the order of the system size, $\ell\sim 1$, meaning that a typical link intersects $c_\T{l}\sim D b^{-1}$ tiles and a typical tile is intersected by $c_\T{t}\sim MD b^{D-1}$ links.
Therefore, in the $N\rightarrow \infty$ large network limit, if we fix the average degree of $\set G$ as $\avg d=2M/N$ and the tile density as $\rho=N/b^{-D}$, the intersection graph has diverging average degrees $c_\T{l}\sim N^{1/D}$ and $c_\T{t}\sim N^{1/D}$ (Fig.~\ref{fig:setup}d).
On the other extreme, if the physical network is lattice-like, i.e., nodes connect to their immediate spatial neighborhood and $\ell\sim b$, the average degrees $c_\T{l}$ and $c_\T{t}$ remain constant.
The diverging average degree $c_\T{t}$ indicates that randomly embedded networks are more susceptible to physical damage than lattice-like networks.
In the following, we systematically investigate the role of physicality in random and targeted damage by relying on $\set I$ both numerically and analytically.

\section{Random damage}
\label{sec:random}

\begin{figure}[b]
	\centering
	\includegraphics[width=1.\columnwidth]{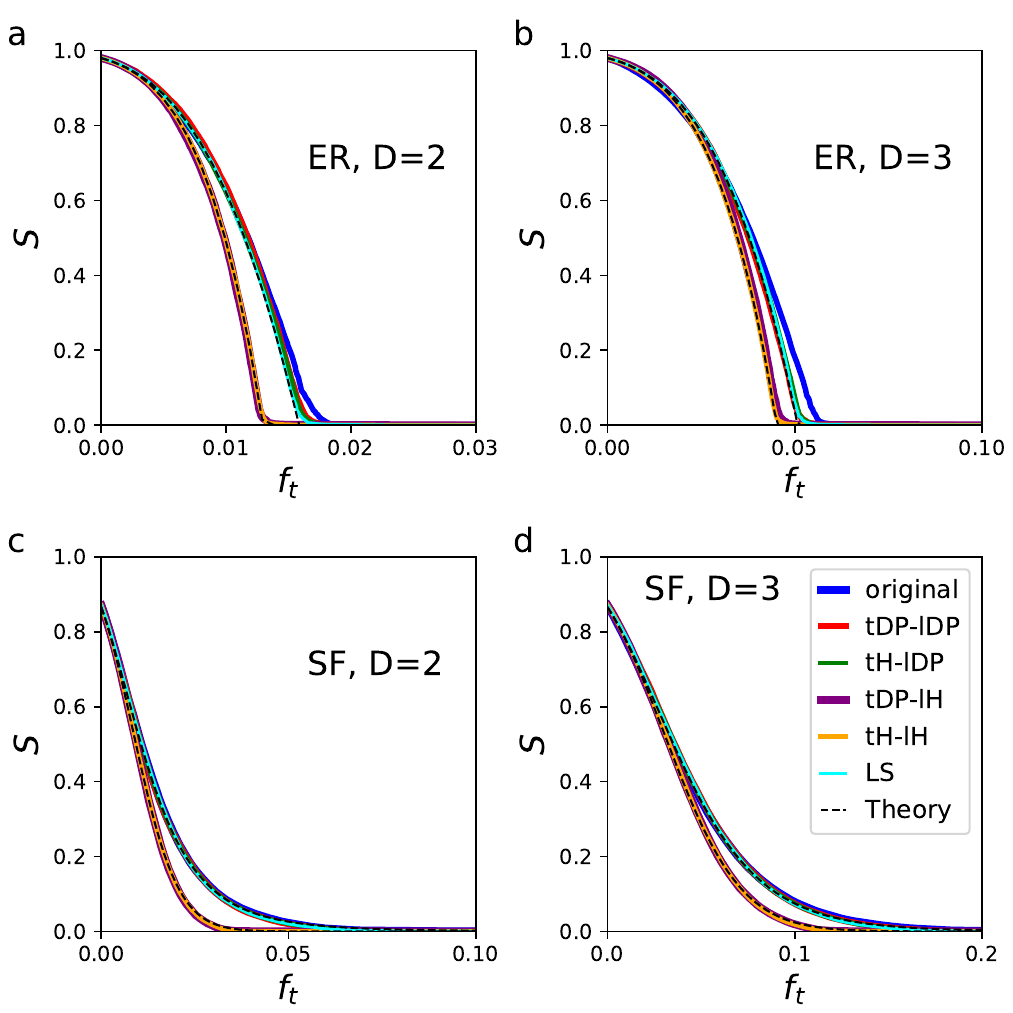}
	\caption{{ \bf Random tile damage in randomly embedded networks.}
	Relative size, $S$, of the largest connected component as a function of the fraction of tiles removed for randomly embedded (a)~Erd\H{o}s-R\'enyi (ER) networks in $D=2$, (b)~ER networks in $D=3$, (c)~scale-free (SF) networks in $D=2$, and (d)~SF networks in $D=3$ dimensions.
	The dashed line represents the theoretical prediction obtained by solving Eqs.~\eqref{eq:s} and \eqref{eq:S}.
	Networks were generated with $N=10^5$ nodes and average degree $c=4$; for SF networks the degree exponent is $\gamma=2.5$, and the tile density is set to $\rho=4$. Each curve is an average of 10 independent realizations.
	\label{fig:rand-S-ft}}
\end{figure}

In this section, we explore the effect of the combinatorial network $\set G$ and the layout $\set P$ on the physical percolation of randomly embedded networks.
Figure~\ref{fig:rand-S-ft} shows the relative size $S$ of the largest component during random tile removal for randomly embedded Erd\H{o}s-R\'enyi (ER) networks and scale-free (SF) networks generated by the static model~\cite{goh2001universal,chung2002connected}. 
We recall that the static model generates networks by fixing the expected degree of each node and allows controlling both the degree exponent $\gamma$ and the average degree of the network $c$. 
We compare $S$ of the original $\set I$ to the five randomized null-models: protocols tDP-lDP, tH-lDP and LS that preserve $P_\T l(k)$ follow the original percolation transition ($S\approx S_\T{tDP-lDP}= S_\T{tH-lDP}= S_\T{LS}$), while link-homogenized randomizations tDP-lH and tH-lH accelerate the transition, shifting $f_\T t ^*$ to the left ($S_\T{tDP-lH}= S_\T{tH-lH}< S$).
This means that, when tiles are damaged independently, only the heterogeneity of the link lengths affects robustness, while tile heterogeneity and correlations between adjacent tiles have negligible effects.
To explain these observations and to systematically investigate random damage, we first develop an analytical solution  for the relative size of the giant component.
Then we use this analytical characterization together with the numerical randomizations of the intersection graph to explore the role of the degree distribution in randomly embedded networks.

\subsection{Analytical characterization}

We start by calculating $S$ for the case when $\set G$ is generated using the configuration model and tiles are removed in random order.
We rely on the standard generating function formalism with the addition that link removal depends on the intersection graph $\set I$, leading to an analytical characterization similar to the so-called feature-enriched percolation~\cite{artime2021percolation}.
In our solution, we assume no correlations between the removed links, an assumption that is generally valid only for the randomized versions of the intersection graph.
We focus on the sparse large network limit $N\rightarrow \infty$, where the average degree of the combinatorial network $c$ remains constant, and we also fix the tile density $\rho=N/b^{-D}$, i.e., the expected number of physical nodes contained in a tile.
This means that the total number of tiles scales as
\begin{equation}
n_\T t \sim N,
\end{equation}
and the size of tiles scales as 
\begin{equation}
b\sim n_\T t^{-1/D}\sim N^{-1/D}.
\end{equation}

Link $e$ is removed if we damage any of its neighbors in $\set I$; therefore, the probability that $e$ is removed after independently damaging $K$ random tiles is
\begin{equation}\label{eq:link_remove_prob}
\begin{aligned}
f(k(e)) &\,= 1 - \left(1-\frac{K}{n_\T t}\right)^{k(e)}\\ 
&\, =1 - \left(1-f_\T t \right)^{k(e)}\approx 1-e^{-k(e) f_\T t},
\end{aligned}
\end{equation}
where $f_\T t=K/n_\T{t}$ is the fraction of tiles removed, and the exponential approximation assumes large $k(e)$.
For straight links, Eq.~\eqref{eq:link_degree} connects the degree of links with its length as $k(e) = 1 + b^{-1}l$, which means that the survival probability of a link drops exponentially with $l$, making long links extremely vulnerable to random tile damage.

Let $s(k)$ denote the probability that a random link with link-side degree $k$ leads to the giant component in $\set G$.
Assuming no correlation between the degree of a link in $\set I$ and its position in $\set G$, we write the self-consistent equation for $s(k)$
\begin{widetext}
\begin{equation}\label{eq:sk}
1 - s(k)= f(k) + (1-f(k))\sum_d q(d)\left(\sum_{k_i}P_\T l(k_i)\left(1-s(k_i)\right) \right)^d,
\end{equation}
\end{widetext}
where $q(d)=(d+1)/c p(d+1)$ is the excess degree distribution of $\set G$. 
Averaging Eq.~\eqref{eq:sk} over $k$, we obtain
\begin{equation}\label{eq:s}
s = G_\T l\left(1-f_\T t\right)\left[1-H\left(1-s\right)\right],
\end{equation}
where $H(z)=\sum_d q(d)z^d$ is the generating function of $q(d)$, and $G_\T l(z) = \sum_k P_\T{l}(k) z^k$ is the moment generating function of the link-side degree distribution of $\set I$.
Similarly, we can obtain the relative size of the giant component $S$ as
\begin{equation}\label{eq:S}
S = 1- G(1-s),
\end{equation}
where $G(z)=\sum_d q(d)z^d$ is the generating function of the degree distribution $p(d)$ of $\set G$.

A crucial quantity in Eq.~\eqref{eq:s} is $1-G_\T l\left(1-f_\T t\right)\equiv f$ providing the probability that a random link is removed, which we calculate by averaging Eq.~\eqref{eq:link_remove_prob} over $P_\T{l}(k)$.
Equivalently, for straight links, we can leverage Eq.~\eqref{eq:link_degree} connecting the degree of a link in $\set I$ to its length to obtain $f$ by averaging over the link lengths, i.e.,
\begin{equation}\label{eq:link_removal_prob_MD}
1-f=\int dl p(l)\left(1-f_\T t \right)^{1+l/b}\approx\int dl p(l)e^{-(1+l/b) f_\T t},
\end{equation}
where $l$ is the link length measured by its Manhattan distance and $p(l)$ is the link length distribution.
Equation~\eqref{eq:link_removal_prob_MD} is valid for the original $\set I$ and the randomizations that do not modify the link-side degree (tDP-lDP and tH-lDP).
On the other hand, randomizations tDP-lH and tH-lH homogenize the link degree in $\set I$, i.e., $k=1+b^{-1}\avg l$ for all links in $\set I_\T r$, where $\avg l$ is the average Manhattan link length.
Therefore, tDP-lH and tH-lH also modify the link removal probability as
\begin{equation}\label{eq:link_removal_prob_lH}
1-f_\T{lH} = e^{-(1+\avg l/b)f_\T t}.
\end{equation}
Invoking Jensen's inequality, we find that
\begin{equation}
f(f_\T t)\leq f_\T{lH}(f_\T t),
\end{equation}
for any $p(l)$, meaning that link length heterogeneity always decreases the number of links removed for a given $f_\T t$~(for example, see Fig.~\ref{fig:rand-S-ft}).

Note that Eqs.~\eqref{eq:s} and \eqref{eq:S} are exactly the equations describing bond percolation in the configuration model with link removal probability $f=1-G_\T l\left(1-f_\T t\right)$.
That is when Eqs.~\eqref{eq:s} and \eqref{eq:S} hold, the only relevant feature of the layout is the link-side degree distribution $P_\T l(k)$ of $\set I$, which affects the percolation transition only through a number of links removed.
The main assumptions we made when deriving these equations is that (i)~$\set G$ is generated by the configuration model, (ii)~the links removed from $\set G$ are uncorrelated, and (iii)~there are no correlations between a link's degree in $\set I$ and it's importance in $\set G$.

\subsection{Effect of degree distribution}
\label{sec:random:degree_dist}

To investigate the role of the degree distribution of the combinatorial network $\set G$, we focus on purely random embeddings.
We select the position of physical nodes uniformly at random from the $D$-dimensional unit cube, $\mathcal{B}_D=[0,1]^{\times D}$, and connect pairs of nodes with straight links.
The Manhattan length of a link $e$ is $l(e) = \sum \lvert r_i - s_i\rvert$, where $\V r\in\mathcal{B}_D$ and $\V s\in\mathcal{B}_D$ are the endpoints of $e$, since $r_i$ and $s_i$ are chosen uniformly from the unit interval, their difference $x=\lvert r_i - s_i\rvert$ follows the distribution $p(x)=2(1-x)$.
Hence, following Eq.~\eqref{eq:link_removal_prob_MD}, we get that the average link removal probability is
\begin{equation}
\begin{split}\label{eq:f_rand_embed}
f =&\, 1- (1-f_\T t)\left(2\int_0^1 dx (1-x) (1-f_\T t)^{x/b}\right)^D \approx\\
\approx &\, 1- e^{-f_\T t}\left(2\int_0^1 dx (1-x) e^{-x f_\T t /b}\right)^D =\\
=&\, 1-e^{-f_\T t}\left(2\frac{e^{-b^{-1}f_t}-1+b^{-1}f_t}{b^{-2}f_\T t^2}\right)^D,
\end{split}
\end{equation}
which is valid for the original $\set I$ and the randomizations that do not modify the link-side degree (tDP-lDP and tH-lDP).
For randomizations tDP-lH and tH-lH that homogenize the link degree, the link removal probability is obtained using Eq.~\eqref{eq:link_removal_prob_lH}, providing
\begin{equation}\label{eq:f_rand_embed_lH}
f_\T{lH} \approx e^{-(1+\avg l/b)f_\T t}=e^{-f_\T t} e^{-\left(1+\frac{D f_\T t}{3b}\right)}.
\end{equation}
In the $N\rightarrow \infty$ large network limit, keeping the tile density $\rho=N/b^{-D}$ constant makes the side length of the tiles to scale as $b\sim n_\T t^{-1/D} \rightarrow 0 $.
Hence the majority of links are removed after damaging only a vanishing fraction of the tiles for both the original and the homogenized distributions $P_\T l(k) $.
Notice, in fact, that the second term of both Eqs.~\eqref{eq:f_rand_embed} and \eqref{eq:f_rand_embed_lH} indicates that almost all links are removed on the scale of $f_\T t\sim n_\T t^{-1/D}$, while the first term for vanishing $f_\T t$ is $e^{-f_\T t}\approx 1$.
This means that the number of tiles to destroy a positive fraction of links in a randomly embedded network scales as $\sim n_\T t^{(D-1)/D}$, i.e., it is on the order of the number of tiles needed to be damaged to cut the network into two parts.

Figure~\ref{fig:rand-S-ft} compares numerical simulations with the analytical estimate of $S$ obtained by inserting Eqs.~\eqref{eq:f_rand_embed} and \eqref{eq:f_rand_embed_lH} into Eq.~\eqref{eq:s}.
We find that the analytical solutions align perfectly with the randomizations and closely follow the simulations using the original $\set I$.
We observe small deviations between the predicted $S$ and original simulated $S$, particularly for ER networks in $D=2$.
To explain this, recall that our calculations assumed that the links are removed from $\set G$ in an uncorrelated fashion. 
This assumption only holds approximately for the original $\set I$: links connected to the same node $v$ tend to intersect the same tiles near the vicinity of $v$, hence having a higher chance of getting removed together.
The effect of such correlations depends on the typical link length $\ell$ and tile size $b$, if $\ell\gg b$ the likelihood of removing a link near its endpoint diminishes.
In the limit $N\rightarrow\infty$ with fixed tile density $\rho$, the relevant length scales are $\ell\sim 1$ and $b\sim N^{-1/D}$; therefore we expect that the above analytical solution correctly captures $S$ for large networks.

\begin{figure}[h]
	\centering
	\includegraphics[width=1.\columnwidth]{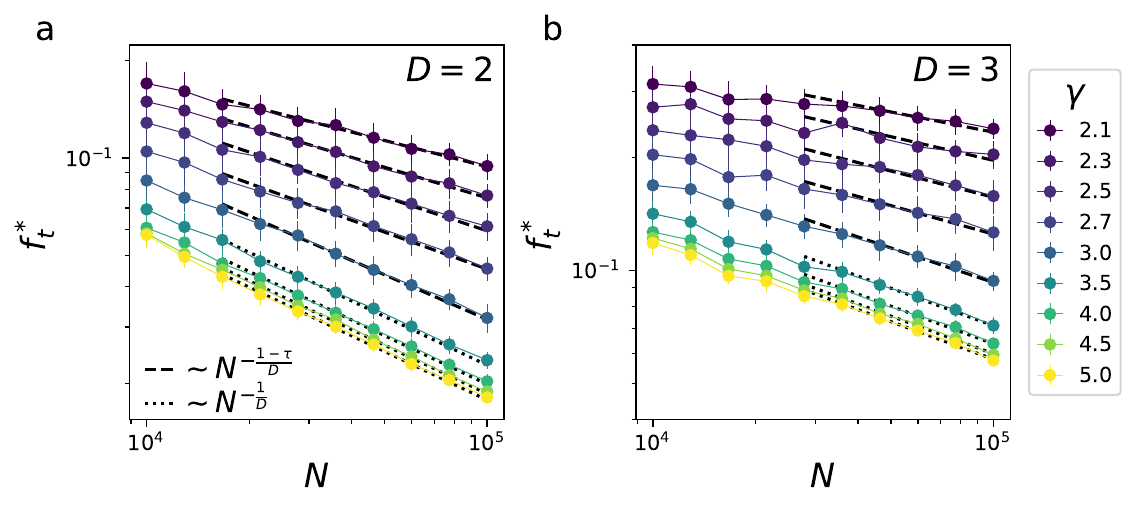}
	\caption{{ \bf The scaling of the critical point for scale-free networks.}
	We simulate random tile damage on randomly embedded SF networks with increasing size $N$ and fixed tile density $\rho$, and we measure the location of the critical point $f_\T t ^*$ by finding the maximum of the second largest component.
	We show the scaling of $f_\T t ^*$ with $N$ for (a)~$D=2$ and (b)~$D=3$ dimensions: markers represent numerical simulations, dashed lines represent the scaling $\sim N^{-(\tau-1)/D}$ predicted by Eq.~\eqref{eq:ft*_sf} for networks with diverging $\avg{d^2}$, and the dotted lines represent the scaling $~N^{-1/D}$ predicted for networks with finite $\avg{ d^2}$.
	We generated networks using $c=4$ and $\rho=4$, the markers represent the average of 50 independent runs, and the error bars represent their standard deviation.
	\label{fig:rand-ft*}}
\end{figure}

The location of the critical point where the giant component is destroyed is often used to quantify the robustness of networks.
To determine the critical point $f^*_\T t$ in our setup, we insert Eq.~\eqref{eq:f_rand_embed} into the Molloy-Reed criterion
\begin{equation}\label{eq:molloy-reed_rand_emb}
\left(2\frac{e^{-b^{-1}f^*_\T t}-1+b^{-1}f^*_\T t}{\left(b^{-1}f^*_\T t\right)^2}\right)^D = \frac{1}{\avg {d^2}/\avg d -1},
\end{equation}
where the left hand side only depends on the physical layout, while the right-hand side only depends on the abstract network structure.
For combinatorial networks with finite second moment $\avg{d^2}$, the right-hand side does not depend on $N$ and hence on $b$, while the left-hand side only depends on $b^{-1}f_\T t = N t^{1/D}f_\T t$. Therefore the critical point vanishes as
\begin{equation}
f_\T{t}^*\sim N^{-1/D}
\end{equation}
in the large network limit.

Scale-free combinatorial networks with $\gamma<3$ have diverging second moments, hence almost all links need to be removed to destroy the giant component in random fashion in the $N\rightarrow\infty$ large network limit, i.e., $f^*\approx 1$.
For traditional random percolation, the finite size scaling of the critical point is characterized by the finite-size scaling $1-f^*\sim N^{-\tau}$, where the exact value of the critical exponent $\tau$ depends on the subtleties of how the scale-free networks are generated~\cite{cohen2002percolation, dorogovtsev2008critical, bhamidi2021multiscale}.
Inserting the latter into Eq.~\eqref{eq:molloy-reed_rand_emb} and keeping only leading terms in $b^{-1}f_\T t$ yields
\begin{equation}
(b^{-1}f_\T t)^{-D} \sim N^{-\tau}, 
\end{equation}
which, in turn, leads to the finite-size scaling
\begin{equation}\label{eq:ft*_sf}
f_\T t^*(\gamma) \sim N^{-\frac{\tau-1}{D}},
\end{equation}
for $3<\gamma<2$. 
To test the above relation, we first numerically estimate the critical exponent $\tau$ by simulating traditional bond percolation on SF combinatorial networks and insert these numerical estimates into Eq.~\eqref{eq:ft*_sf}.
Figure~\ref{fig:rand-ft*} compares the theoretically obtained scaling relation to numerical simulations of $f_\T t^*$ for physically embedded SF networks, finding a good agreement for large $N$.

Overall, we demonstrated that randomly removing a vanishingly small fraction of tiles is sufficient to dismantle randomly embedded networks and that higher dimensional embeddings increase their vulnerability.
This is because of the presence of links whose length scales with the linear size of the unit cube, i.e.\ $\ell\sim1$, even in the $N\rightarrow\infty$ limit.
%with typical length $\ell\sim 1$, these links do not scale with the network size $N$, i.e., their typical length remains  $\ell\sim 1$ in the $N\rightarrow\infty$ limit.
Therefore, the number of tiles the links intersect diverges as $\sim b^{-1}\sim N^{1/D}$, thus making them extremely vulnerable to physical damage.
We also found that the analytical solution, which in general is valid for the randomized versions of the intersection graph, well describes the percolation transition for the original $\set I$ in large networks, revealing that random tile damage is equivalent to a random bond percolation, where the number of links we remove is determined by the layout $\set P$. 

\section{Targeted damage}
\label{sec:targeted}

We now turn our attention to targeted physical attacks, where we iteratively damage a fraction $f_t$ of the tiles having the highest degree in $\set I$.
More formally, given a tiling described by the intersection graph $\set I$:
\begin{enumerate}
\item we find tile $t$ with the highest degree in $\set I$;
\item we remove $t$ from $\set I$ together with each of its neighbors, i.e., with each link $e$ that intersects $t$;
\item we repeat steps 1 and 2 until all links are removed.
\end{enumerate}

\begin{figure}[h]
	\centering
	\includegraphics[width=1.\columnwidth]{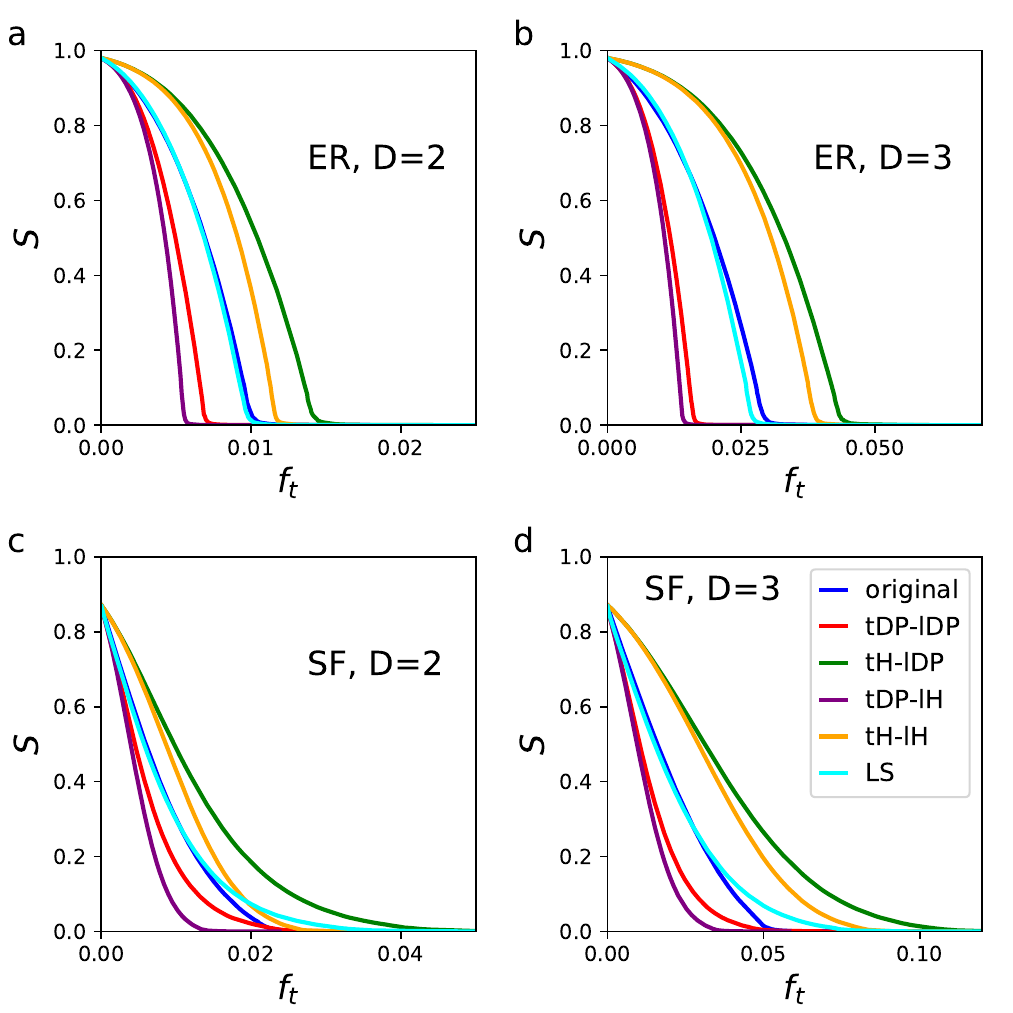}
	\caption{{ \bf Targeted attack of randomly embedded networks.}
	We show the relative size $S$ of the largest connected component as a function of the fraction of tiles removed for randomly embedded (a)~Erd\H{o}s-R\'enyi (ER) networks in $D=2$, (b)~ER networks in $D=3$, (c)~scale-free (SF) networks in $D=2$, and (d)~SF networks in $D=3$ dimensions.
	Networks were generated with $N=10^5$ and $c=4$, for SF networks the degree exponent is $\gamma=2.5$, and the tile density is set to $\rho=4$. Each curve is an average of 10 independent realizations.
	\label{fig:att-S-ft}}
\end{figure}

As before, we investigate the role of the degree distribution of $\set G$ and explore the simpler case of randomly embedded networks.
Figures~\ref{fig:att-S-ft}a,~b show the size of the largest component of a randomly embedded Erd\H{o}s-R\'enyi (ER) network in $D=2$ and $D=3$ dimensions under targeted physical attacks.

In Sec.~\ref{sec:random}, we found that the only relevant property affecting $S$ during random tile removal is the link-side degree distribution of $\set I$. 
In the case of targeted physical damage, a richer picture emerges. 
By comparing the original $\set I$ with its randomized versions, we find that $S_\T{tDP-lH}<S_\T{tDP-lDP}<S\approx S_\T{LS}<S_\T{tH-lH}<S_\T{tH-lDP}$ for the entire range of $f_\T t$.
This means that (i)~tile heterogeneity increases the vulnerability of the network, (ii)~correlations in $\set I$ (such as short loops caused by neighboring links intersecting the same tiles), and link heterogeneity makes the network more robust against targeted attacks.
Finally, we find that $S$ for the original $\set I$ is identical to the case where we randomly remove the same number of links (LS randomization), meaning that for the ER network, targeted attacks dismantle the network faster than random tile removal, because it damages more links per tile, but not remove links that are important for network cohesion. 

\begin{figure*}[t!]
	\centering
	\includegraphics[width=1.5\columnwidth]{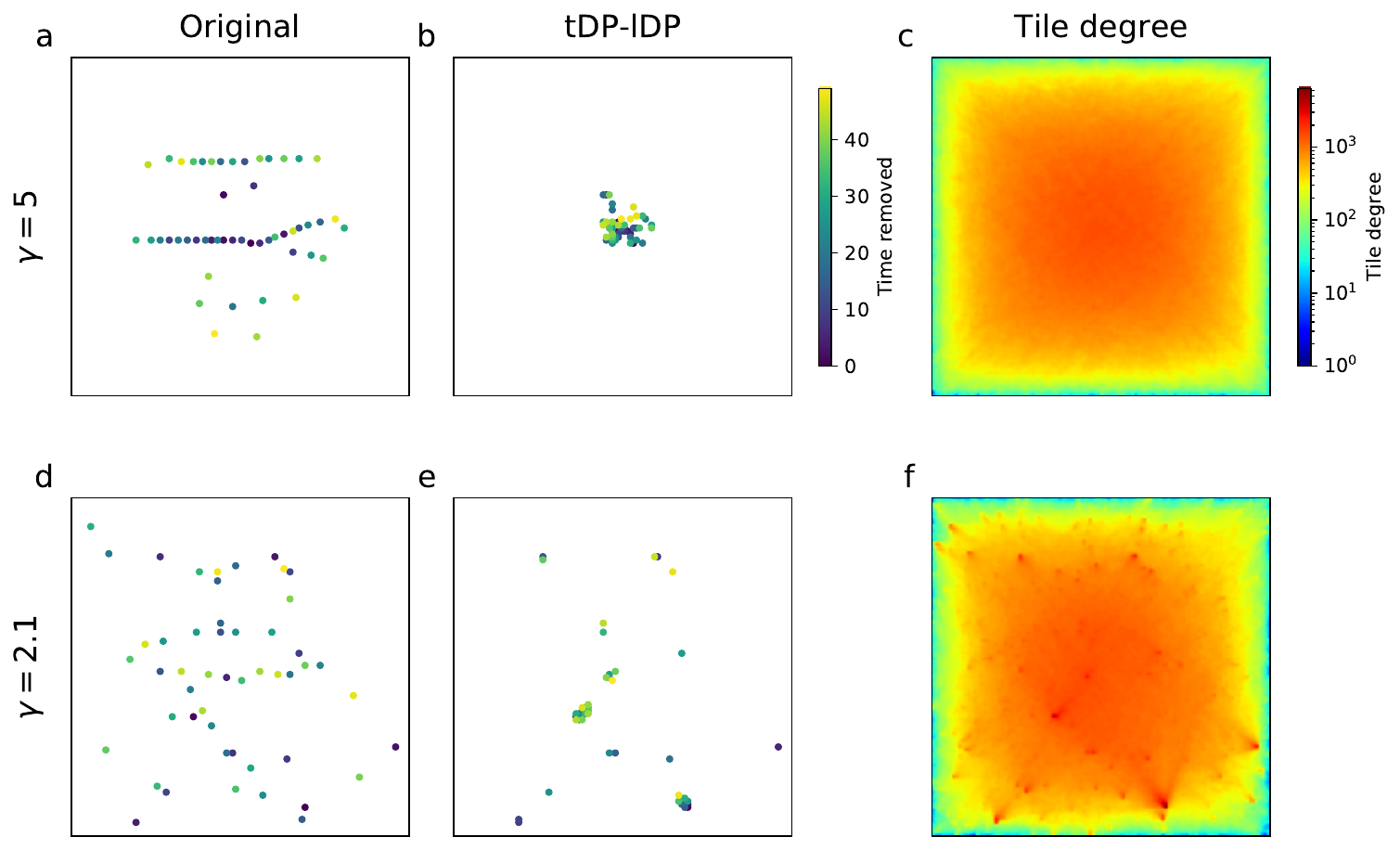}
	\caption{{ \bf Example targeted tile removal.}
	(a)~The location of the first 50 tiles removed from a SF network with degree exponent $\gamma=5$. The removed tiles
	(b)~The tDP-lDP randomization of the network removes the overlap between the links intersected by neighboring tiles, therefore, the attack targets the center tiles.
	(c)~The spatial distribution of the tile-degree shows that the center tiles intersect the most links, and monotonically drops towards the edges of the square.
	(d)~The first 50 tiles removed from a SF network with degree exponent $\gamma=2.1$ target the hubs in $\set G$.
	(e)~In the case of the tDP-lDP randomization, tile removal targets the tiles surrounding the highest degree nodes.
	(f)~The spatial distribution of the tile-degree has peaks at tiles that contain hubs of $\set G$.
	The figure shows single networks with $N=5\times 10^4$, $c=4$, and $\rho=4$, the color of the tiles in (a), (b), (d), and (e) indicates the order that the tiles are removed in.
	\label{fig:att-example}}
\end{figure*}

Figures~\ref{fig:att-S-ft}c,~d show instead the relative size of the giant component, $S$, for SF networks with degree exponent $\gamma = 2.5$, thus having a divergent second moment. 
The portrait, in this case, is analogous to the one observed in ER networks, with one key exception: while initially $S\approx S_\T{LS}$, near the critical point the original network falls apart faster than the LS and even the tDP-lDP version, i.e., $f_\T{t}^*<f_\T{t, tDP-lDP}^*<f_\T{t, LS}^*$.
This means that correlations between $\set I$ and $\set G$ accelerate the targeted disruption of the network.

To further clarify the role of degree heterogeneity in $\set G$, we plot the location of the initial 50 tiles removed for a $D=2$ SF network with $\gamma = 5$ and with $\gamma = 2.1$, for both the original network (Fig.~\ref{fig:att-example}a and d) and the tDP-lDP randomization (Fig.~\ref{fig:att-example}b and e).
For the original homogeneous network ($\gamma=5$), the initial removal of the tiles concentrates in the center of the embedding square, forming a spatial structure that cuts through the network along a straight line (for $D=3$ such a structure is less apparent).
In contrast, for the heterogeneous network ($\gamma=2.1$), the removed tiles are randomly scattered in space.

To understand this pattern, consider that the spatial distribution of links in homogeneous networks is approximately the same as picking uniformly random segments from the unit square.
As a result, we expect tile degree $k(t)$ to be the highest in the center of the embedding square and monotonically decreasing toward the edges (Fig.~\ref{fig:att-example}c), explaining why the center tiles are picked initially.
However, the highest-degree center tiles are redundant: they are intersected by the same links, i.e., they share many neighbors in the intersection graph $\set I$.
The one-dimensional structure of the removed tiles emerges as a result of the targeted attack greedily minimizing the redundancy between the removed tiles.
Indeed, the tDP-lDP randomization keeps the degree of the tiles but removes correlations between them, hence the initial removed tiles concentrate at the center of the embedding square forming a two-dimensional patch (Fig.~\ref{fig:att-example}b).
This also means that the total number of links removed by the tDP-lDP randomization increases compared to the number of links removed by the original $\set I$, explaining the observation $S_\T{tDP-lDP}<S$. 

For heterogeneous networks, on the other hand, links are not uniformly distributed: tiles that contain the hubs of $\set G$ intersect an out-sized number of links, resulting in an uneven spatial distribution of the tile-degree (Fig.~\ref{fig:att-example}f).
The random distribution of the initially removed tiles is a result of the targeted attack selecting tiles that contain the hubs of $\set G$.
The tDP-lDP randomization removes redundancy between the tiles surrounding the hubs, hence tile removal is concentrated on the few largest hubs of the network, explaining why $S_\T{tDP-lDP}<S$ in the early stages of the process.
The LS and tDP-lDP randomizations remove links in an uncorrelated way from $\set G$, meaning that hubs are not removed in a single step but lose instead links in a continuous fashion.
The fact that tile removal for LS and tDP-lDP does not eliminate hubs entirely, thus delaying the critical point of the transition, explains the observed $f_\T{t}^*<f_\T{t, tDP-lDP}^*<f_\T{t, LS}^*$.

\begin{figure}
	\centering
	\includegraphics[width=1.\columnwidth]{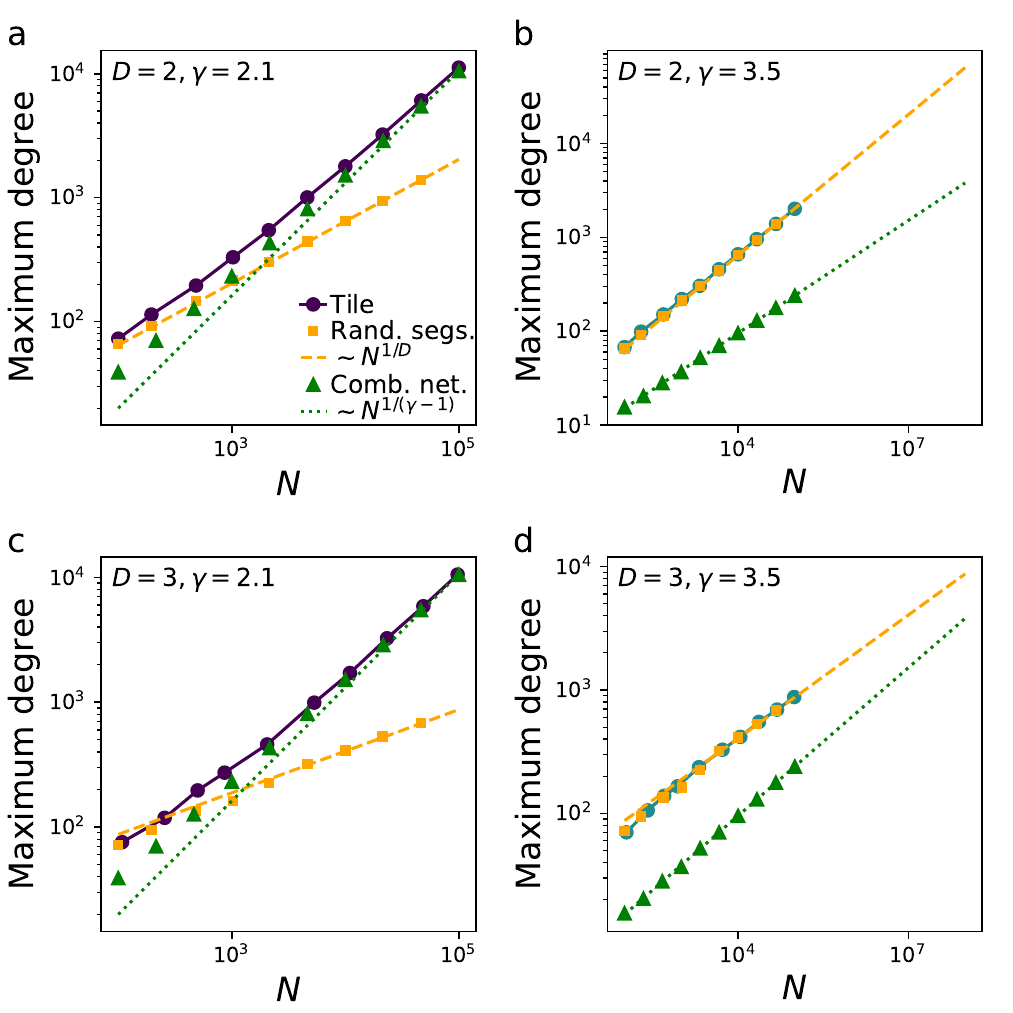}
	\caption{{ \bf Scaling of the maximum tile-degree.}
	We measured the maximum tile degree (circles) for increasing size $N$ for SF networks embedded in $D=2$ and $D=3$ dimensions and with $\gamma=2.1$ and $\gamma=3.5$ degree exponents and average degree $c=4$.
	We randomly placed $cN/2$ segments in the unit square and measured the maximum number of segments intersecting a tile (circles), and we also measured the maximum degree of the combinatorial networks (triangles).
	The dashed lines represent the scaling predicted by Eqs.~\eqref{eq:k_pass} and \eqref{eq:k_nodes}. 
	(a,b)~For $\gamma=2.1$, the contribution of the hubs dominates the maximum tile degree for both $D=2$ and $D=3$.
	(c)~For $\gamma=3.5$ and $D=2$ dimensions, $\gamma>D+1$, hence the contribution of $\max k_\T{pass}$ dominates.
	(d)~In contrast, for $\gamma=3.5$ and $D=3$, we predict that $\max k_\T{nodes}$ dominate. However, extrapolating the simulations shows that $\max k_\T{nodes}$ only outgrows $\max k_\T{pass}$ for extremely large networks.
	Circle markers represent the average of 20 independent physically embedded networks, while triangles and squares represent the average of 100 runs.
	\label{fig:att-scaling}}
\end{figure}

In the above we have shown that, for homogeneous networks, tiles with the most links passing through get targeted first, while, for heterogeneous networks, the location of the hubs of $\set G$ determines which tiles get removed.
To explore when node degree dominates tile removal, divide degree $k(t)$ of tile $t$ in $\set I$ into two contributions: 
\begin{equation}
k(t) = k_\T{pass}(t) + k_\T{nodes}(t),
\end{equation}
where $k_\T{pass}(t)$ is the number of links that pass through $t$, i.e., links that intersect $t$, but have endpoints outside $t$, and $k_\T{nodes}(t)$ is the number of links that have at least one endpoint in $t$. 
We estimate the maximum of the two contributions in the $N\rightarrow \infty$ large network limit with fixed tile density $\rho=Nb^D$.
First, note that the tile $t_0$ at the center of the unit square or cube has the highest expected number of links passing through, i.e., links that intersect $t_0$ but have endpoints outside of $t_0$.
The probability that a randomly placed segment $s$ intersects $t_0$ is proportional to the cross section of $t_0$ perpendicular to $s$, which is $\sim b^{D-1}\sim N^{-\frac{D-1}{D}}$.
The number of links in the network is $\sim N$; therefore, the total number of links passing through $t_0$ scales as
\begin{equation}\label{eq:k_pass}
\max \left[k_\T{pass}(t)\right]\sim N\cdot N^{-\frac{D-1}{D}}=N^{\frac{1}{D}}.
\end{equation}
The scaling of the maximum contribution of $k_\T{pass}(t)$ is provided by scaling the largest degree node in $\set G$, hence it depends on the topology of the network. For heterogeneous networks generated in the static model~\cite{lee2006intrinsic}, this is given by $k_\T{max}\propto N^\theta$ with $\theta=1/(\gamma-1)$, so that
\begin{equation}\label{eq:k_nodes}
\max \left[k_\T{nodes}(t)\right]\sim N^{\frac{1}{\gamma-1}}.
\end{equation}
The above two scaling relations determine whether the center tile or the tile containing the largest hub of the network gets removed in the first step of the targeted tile removal: for $\gamma<D+1$, the value of $\max\left[ k_\T{nodes}\right]$ outgrows $\max \left[k_\T{pass}\right]$ and initially the hubs get removed, otherwise the targeted attack removes the tiles close to the center.
Figure~\ref{fig:att-scaling} illustrates the scaling of the maximum tile degree.
For $\gamma=2.1$, the contribution of hubs $\max\left[ k_\T{nodes}\right]$ dominates both in $D=2$ and $D=3$ dimensions.
In contrast, for $\gamma=3.5$, the contribution $\max\left[ k_\T{pass}\right]$ dominates for $D=2$, while for $D=3$, hubs eventually outgrow the effect of links passing through.
Note, however, that the latter only happens for very large networks outside of the regime of most real networks.

Before exploring the effect of physical damage in real datasets, let us stress that the geometric effects detailed above can depend on the subtleties of how the networks are generated. In the static model for scale-free networks, for example, one has the natural degree cut-off $k_{max}\propto N^\theta$, where $\theta=1/(\gamma-1)$, above which degree correlations start to appear. Similarly, for randomly growing SF networks, one has $\theta=1/(\gamma-1)$, while in the uncorrelated configurational model~\cite{dorogovtsev2008critical}, $\theta$ generally depends on multi-degree or single-edges constraints and it is, respectively, given by $\theta=1/(\gamma-1)$ and $\theta=1/2$~\cite{bhamidi2021multiscale}. The latter scaling, in particular, suggests that in the targeted removal of tiles, hubs and center tiles are removed equally likely in the early stages only for $D=2$, while hubs are always the first ones to be removed for embedding dimensions $D\geq3$.

\section{Empirical networks}
\label{sec:empirical}

In this section, we analyze the robustness of three empirical networks using the intersection graph: an airline network, a vascular network and a neural network.
We perform both random and targeted tile removal, and compare the percolation transition obtained for the original layout against the randomized null models.
Finally, we test the robustness of our results by systematically changing the size of the tiles. 

A difference between model networks and empirical data is that we embedded model networks in the unit square or cube. Real networks, on the other hand, typically have a less regular shape, hence their bounding box may contain large empty regions.
Therefore, to tile an empirical network, we identify its axis-aligned bounding box, we tile this bounding box with cubic boxes, and, crucially, we leave out the empty tiles from our analysis.
In other words, we remove isolated vertices from the intersection graph $\set I$ before simulating tile removal.
In the following, we discuss each case separately.

\subsection{Air traffic network}

\begin{figure}[!t]
	\centering
	\includegraphics[width=1.\columnwidth]{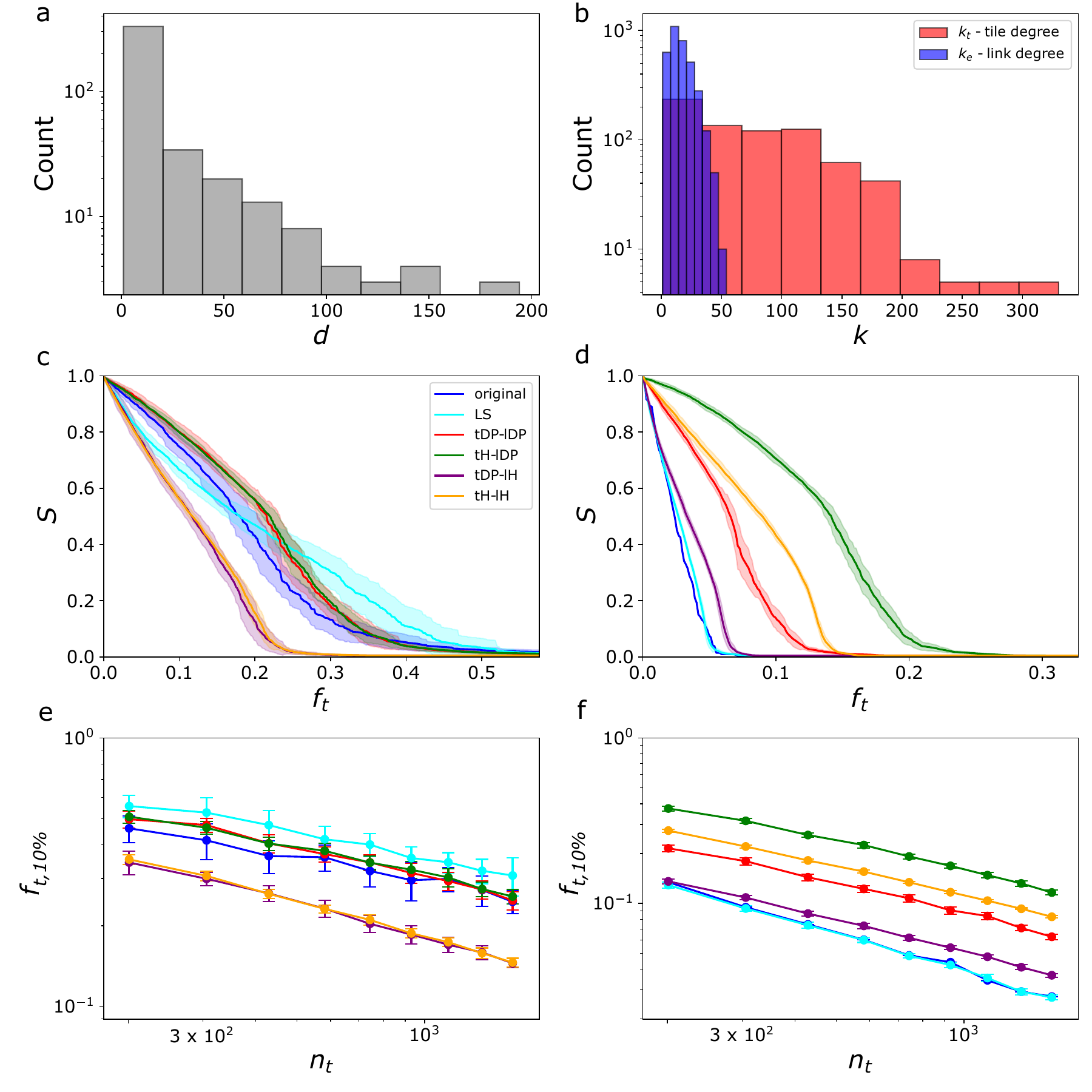}
	\caption{{ \bf Air traffic network.}
	(a)~The degree distribution $p(d)$ of the combinatorial network. The largest hub is connected to 194 of the $N=419$ nodes of the network.
	(b)~The tile-side degree distribution $P_\T{t}(k)$ and the link-side degree distribution $P_\T{e}(k)$ of the intersection graph $\set{I}$.
	(c,d)~The size of the largest component $S$ during random and targeted tile removal.
	(e,f)~The fraction of tiles $f_\T{t,10\%}$ needed to be removed to reduce the largest component to $S=0.1$ for random and targeted tile removal as a function of the number of non-empty tiles $n_\T{t}$ used to cover the network.
	The order of the randomized variants is insensitive to the choice of tile size.
	In panels (c-f), lines and markers represent an average of 20 independent randomizations, and the error bars indicate the standard deviation.}
	\label{fig:airline}
\end{figure}

Our first case study is a network representing air traffic in the contiguous US, which we constructed using data available from the Bureau of Transportation Statistics for the year 2023~\cite{bts}.
The network contains $N=419$ nodes representing cities, and we added a link between two cities if the total number of passengers on direct flights between the pair exceeded $1000$, resulting in an average degree of $c\approx 16.7$.
We obtain the coordinates of the cities from OpenStreetMap and we transform the longitude and latitude pairs to Euclidean space using the Albers equal area projection~\cite{openstreetmap,pyproj}.
We consider flight paths to be straight lines between cities.

Figure~\ref{fig:airline}a shows the degree distribution $p(d)$ of the combinatorial network $\set G$.
We find that $p(d)$ is highly heterogeneous: the median degree is $4$, while the largest hub is connected to 194 nodes, representing close to half of the cities.
To construct the intersection graph $\set{I}$, we tile the network such that we place 40 square tiles along the longest axis of its bounding box, resulting in $n_\T{t} = 743$ non-empty square tiles and tile density $\rho\approx 0.56$.
Flights often traverse the US, hence the longest links in the network are comparable to the size of the bounding box.
As a consequence, the tile-side degree distribution $P_\T{t}(k)$ and the link-side degree distribution $P_\T{l}(k)$ of $\set{I}$ resemble the degree distributions observed in randomly embedded networks (Fig.~\ref{fig:airline}b), where there is also no constraint on the maximum link length.

Figure~\ref{fig:airline}c shows the size of the largest component $S$ as a function of the tiles removed.
Due to the high average degree $c\approx 16.7$, a large fraction of the links are needed to be removed to dismantle the network: for traditional bond percolation, we must remove approximately $f\approx 0.97$ fraction of the links from $\set G$ to reduce its largest component to $S=0.1$.
In stark contrast, the same reduction in $S$ is achieved by randomly removing only $f_\T t \approx 0.24$ fraction of the tiles.
Comparing the original intersection graph $\set I$ to its randomizations, we find that tile-degree heterogeneity by itself has little effect and that the link-side degree heterogeneity delays the percolation ($S_\T{tDP-lH}\approx S_\T{tH-lH}<S_\T{tDP-lDP}\approx S_\T{tH-lDP}$), similarly to randomly embedded model networks (Fig.~\ref{fig:rand-S-ft}).
However, in contrast to randomly embedded networks, we find that the LS randomization initially reduces $S$ faster than the original $\set I$ ($S>S_\T{LS}$), but eventually LS delays the percolation transition ($S<S_\T{LS}$). 
Recall that LS randomization removes the same number of links from $\set G$, but randomly; therefore, the above observations suggest that there is a correlation between a link's degree in $\set I$ and its importance in $\set G$.
Indeed, calculating the Pearson correlation between link-degree and the product of the degree of a link's endpoints in $\set G$, we find a positive correlation of $r\approx 0.26$.
This means that the original tile removal tends to remove links connecting hubs faster than the LS randomization, which explains the observed pattern: removing links between hubs in the network with high $c$ initially does not reduce $S$, but in the long run accelerates the destruction of the largest component.

For targeted tile removal, in Sec.~\ref{sec:targeted}, we found that tiles containing the largest hubs are removed first from degree heterogeneous networks, hence we expect a similar pattern for the airline network.
Indeed, the first five tiles removed, for example, all contain major airline hubs, such as Denver, Dallas, or Atlanta.
Figure~\ref{fig:airline}d compares the original $S$ to its randomized counterparts, and we find a similar pattern to randomly embedded networks: $S\approx S_\T{LS} < S_\T{tDP-lH}<S_\T{tDP-lDP}<S_\T{tH-lH}<S_\T{tH-lDP}$, meaning that tile-degree heterogeneity accelerates, while link-degree heterogeneity slows down the percolation process.
A key difference compared to model networks is that the original removal decays even faster than the tDP-lH and tDP-lDP randomizations.
To explain this observation, note that the links intersecting tiles that contain hubs in $\set I$ have less overlap than in the randomized versions.
For example, in our network Dallas has 194 and Denver has 182 connections, but only one of these connections, namely flights between Dallas and Denver, overlap.
The expected overlap between random sets of links of the same size is $194\cdot 182/3500\approx 10$, hence when removing the tiles containing Dallas and Denver, more unique links are damaged by the original process than by the randomized variants.

Finally, to test the effect of tile size, we measure the fraction of tiles $f_\T{t,10\%}$ needed to be removed to reduce the largest component to $S=0.1$ for tilings of various sizes.
Figure~\ref{fig:airline}e and f show that the order of the randomizations does not change in the entire range of tiles that we tested both for random and targeted percolation.

\subsection{Vascular network}

\begin{figure}[h]
	\centering
	\includegraphics[width=1.\columnwidth]{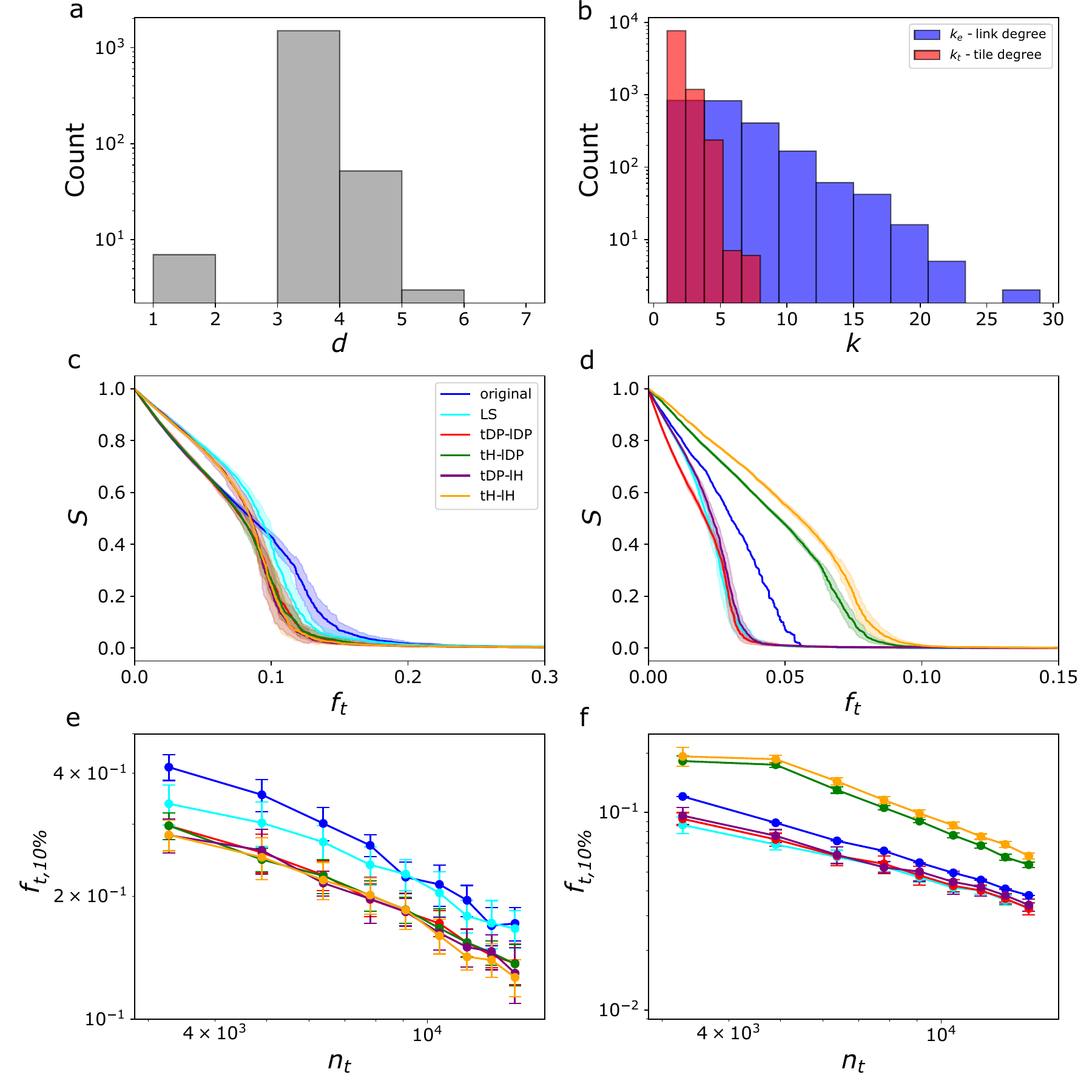}
	\caption{{ \bf Vascular network.}
(a)~The degree distribution $p(d)$ of the combinatorial network. The majority of nodes are bifurcation points with degree $d=3$.
	(b)~The tile-side degree distribution $P_\T{t}(k)$ and the link-side degree distribution $P_\T{e}(k)$ of the intersection graph $\set{I}$.
	(c,d)~The size of the largest component $S$ during random and targeted tile removal.
	(e,f)~The fraction of tiles $f_\T{t,10\%}$ needed to be removed to reduce the largest component to $S=0.1$ for random and targeted tile removal as a function of the number of non-empty tiles $n_\T{t}$ used to cover the network.
	The order of the randomized variants is insensitive to the choice of tile size.
	In panels (c-f), lines and markers represent an average of 20 independent randomizations, the error bars indicate the standard deviation.}
	\label{fig:vascular}
\end{figure}

Our second case study is a network representing the vasculature in a sample of the brain of a mouse~\cite{gagnon2015quantifying}.
In the network, nodes represent branching points of the vessels or terminal points at the edge of the sample, while links represent vessels in between branching point, overall resulting in $N=1558$ nodes and $M=2359$ links.
Note that links are not straight lines, but follow a winding trajectory.
Figure~\ref{fig:vascular}a shows that the degree distribution $p(d)$ highly homogeneous, largely concentrated on $d=3$, which indicates that most branching points split vessels into two new branches.
For more details about the properties of the network see Ref.~\cite{blagojevic2024three}.

We tile the network with cubes such that 20 tiles are placed along the longest axis of the network's bounding box.
After dropping the empty tiles, the network is covered by $n_\T{t} = 3276$ boxes, resulting in a tile density of $\rho \approx 0.48$.
In contrast to the airline network, Fig.~\ref{fig:vascular}b shows that the link-side degree distribution $P_\T{l}(k)$ is peaked at $k=1$, meaning that typical links are short compared to tile size $b$.
As a consequence, the average tile-side degree is also much lower compared to the airline network.
The low link-side degrees of $\set I$ together with the homogeneous degree distribution of $\set G$ makes the vascular network lattice-like, counter to the airline network and the randomly embedded model networks.

For random tile removal, Fig.~\ref{fig:vascular}c shows that the randomizations overlap, especially in the later stages of the percolation process ($S_\T{tDP-lDP} \approx S_\T{tDP-lH} \approx S_\T{tH-lDP} \approx S_\T{tH-lH}$).
This is explained by the fact that $P_\T l(k)$ follows an exponential distribution; therefore further homogenizing it has little affect on random removal.
A curious pattern is that the original process is slower at dismantling the largest component than randomly removing the same number of links ($S<S_\T{LS}$).
To understand this, notice that the majority of links intersect only a few tiles; therefore a link $e$ is likely to be removed at its endpoint node $v$, together with other links adjacent to $v$.
Links connected to the same node $v$ play a redundant role in the connectivity of the network, hence removing them together reduces $S$ slower than removing the same number of random links.

For targeted removal, Fig.~\ref{fig:vascular}d shows that tile-side degree heterogeneity matters ($S_\T{tDP-lDP}\approx S_\T{LS} \approx S_\T{tDP-lH} < S< S_\T{tH-lDP} < S_\T{tH-lH}$), meaning that although $P_\T t(k)$ is exponentially distributed, targeting its tail still gives an advantage.
Similarly to random removal, we observe that $S<S_\T{LS}$, which is again explained by removing links at their endpoint and the fact that $p(d)$ lacks hubs that would need to be removed.

Overall, tH null models are more robust than the original layout, which would make sense as for homogenized tiles, each one would be equally likely, thus resembling the random attack scenario. On the other hand, all other null models are less robust and break down quicker than the original layout. One potential explanation is that even though some regions contain many dense tiles, which might cause redundant removals that effectively just prune the network, instead of severing it into larger disconnected components.

Finally, to test the effect of tile size, we measure the fraction of tiles $f_\T{t,10\%}$ needed to be removed to reduce $S$ to $0.1$.
Similarly to the airline network, Fig.~\ref{fig:vascular} and f show that the order of the randomizations does not change in the entire range of tiles that we tested both for random and targeted percolation.

\subsection{Neural network}
 
\begin{figure}[h]
	\centering
	\includegraphics[width=1.\columnwidth]{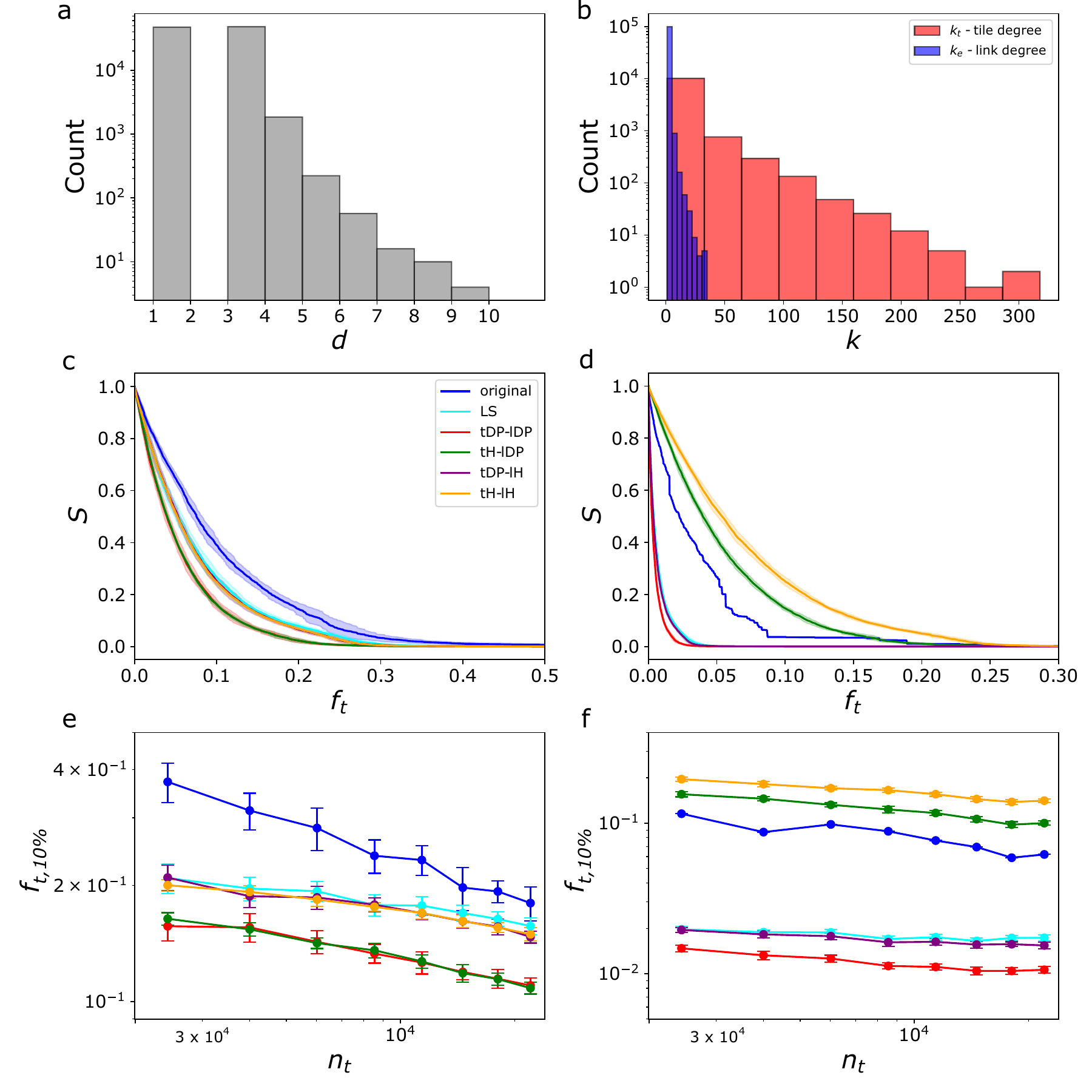}
	\caption{{ \bf Neural network.}
	(a)~The degree distribution $p(d)$ of the combinatorial network. Majority of nodes are terminal or bifurcation points with degree $d=1$ or $d=3$, respectively. The average degree of the network $c\approx 2.06$ is close to two, which would corresponds to a tree.
	(b)~The tile-side degree distribution $P_\T{t}(k)$ and the link-side degree distribution $P_\T{e}(k)$ of the intersection graph $\set{I}$.
	(c,d)~The size of the largest component $S$ during random and targeted tile removal.
	(e,f)~The fraction of tiles $f_\T{t,10\%}$ needed to be removed to reduce the largest component to $S=0.1$ for random and targeted tile removal as a function of the number of non-empty tiles $n_\T{t}$ used to cover the network.
	The order of the randomized variants is insensitive to the choice of tile size.
	In panels (c-f), lines and markers represent an average of 20 independent randomizations, and the error bars indicate the standard deviation.}
	\label{fig:fruitfly}
\end{figure}

As the final case study, we analyze a network of neurons comprising a region of the central nervous systems of a fruit fly~\cite{clements2020neuprint}.
The network is composed of 96 neurons and 6249 synaptic connections.
Each neuron in the data set is represented as a spatially embedded tree and these embedded trees are bound together through synapses. 
Here, we focus on a microscopic representation of this network: we treat branching points and terminal points of the linked trees as nodes, and we treat the connections between them as links.
This way our network contains $N=97,588$ nodes and $M=100,388$ links, making it our largest example.
Figure\ref{fig:fruitfly}a shows that the degree distribution $p(d)$ of $\set G$ is concentrated on $d=1$ and $d=3$, indicating that most nodes are terminal or bifurcation points of the neurons. 
For more details about the properties of this network, see Ref.~\cite{blagojevic2024three}.

We tile the network with cubes such that 40 tiles are placed along the longest axis of the network's bounding box.
After dropping the empty tiles, the network is covered by $n_\T{t} = 11,389$ boxes, resulting in a tile density of $\rho \approx 8.6$.
Figure~\ref{fig:fruitfly}b shows that the link-side degree distribution $P_\T{l}(k)$ is sharply peaked at $k=1$, meaning that typical links are shorter than $b$ and are contained inside a single tile.
Note, however, that despite the peak at $k=1$ there are still a few links that span the bounding box. 
We observe a high maximum tile-side degree, this is due to tiles that contain many nodes.
Although there are some key differences, the high peak of $P_\T l(k)$ at $k=1$ and the homogeneous $p(d)$ make the fruit fly neural network similar to the vascular network.

For random tile removal, Fig.~\ref{fig:fruitfly}c shows a very similar pattern to the vascular network, with the difference that link-degree does have an effect.
For targeted removal, shown in Fig.~\ref{fig:fruitfly}d, the ordering of the randomizations is identical to that of the vascular network.

Finally, as in both cases before,  Fig.~\ref{fig:vascular} and f show that the order of the randomizations is not sensitive to the choice of tile size in the ranges that we tested.

\section{Discussion}

In this article, we have proposed a framework to explore the vulnerability of complex networks against physical damage.
The setup aims to model spatially localized attacks and it takes into account the routing of the links.
One key observation is that long links are necessarily susceptible to physical damage, hence their presence makes networks extremely vulnerable.
Traditional network science focuses solely on combinatorial networks, while spatial network theory also takes into account the coordinates of nodes but in most cases ignores link routing.
Our results highlight that incorporating the physical shape of links and nodes can reveal properties of networks that are otherwise missed by more traditional approaches.
 
The central tool of our analysis is the intersection graph, which concisely captures how the physical layout affects the percolation transition.
Faithfully representing the shape of physical links and nodes requires large amounts of data; therefore directly working with such representations is computationally expensive and makes analytical description difficult.
Calculating the intersection graph provides a way to extract the relevant properties of the physical layout of a network, thus simplifying numerical exploration and enabling us to adapt the analytical tools of network science to characterize the percolation transition. 
The intersection graph, and the conceptually similar meta-graph of Ref.~\cite{posfai2024impact} or contactome of Ref.~\cite{salova2024combined}, may provide a blueprint for tackling the complexity of physical layouts.

Our work raises several new questions.
For example, we studied randomly embedded model networks, which allowed simple analytical characterization.
Future work may extend this to more realistic models which take into account distance when creating links~\cite{barthelemy2011spatial} or other physical constraints~\cite{posfai2024impact,pete2024physical}, or may explore the role of the spatial organization of links, such as bundling~\cite{bonamassa2024bundling} and entanglement~\cite{liu2021isotopy, glover2024measuring}.
Also, throughout this paper, we focused on network embeddings where nodes are point-like and links are extended objects connecting them.
Many physically embedded networks, however, are better characterized by extended nodes with point-like connections between them, e.g., neurons are extended objects with complex shapes, which connect to each other via point-like synapses.
Such physical networks are better represented as spatially embedded network-of-networks and future work may extend the tile removal percolation to such representations~\cite{bianconi2013superconductor,chepuri2023complex,pete2024physical}.

\paragraph*{Acknowledgments.}
This work was supported by ERC grant No.~810115\nobreakdash-DYNASNET.

\bibliographystyle{unsrt}
\bibliography{tile_perc.bib}

\end{document}